# XPS evidence for molecular charge-transfer doping of graphene

Debraj Choudhury[1,2], Barun Das[1], D. D. Sarma[1,*], and C. N. R. Rao[1,3]

[1] Solid State and Structural Chemistry Unit, Indian Institute of Science,
Bangalore-560012, India.
[2] Department of Physics, Indian Institute of Science, Bangalore-560012, India.
[3] New Chemistry Unit and International Centre for Materials Science (ICMS), Jawaharlal Nehru Centre for
Advanced Scientific Research (JNCASR), Jakkur P.O., Bangalore -560 064, India.

## Abstract

By employing x-ray photoelectron spectroscopy (XPS), we have been able to establish the occurrence of charge-transfer doping in few-layer graphene covered with electron acceptor (TCNE) and donor (TTF) molecules. We have performed quantitative estimates of the extent of charge transfer in these complexes and elucidated the origin of unusual shifts of their Raman G bands and explained the differences in the dependence of conductivity on n- and p-doping. The study unravels the cause of the apparent difference between the charge-transfer doping and electrochemical doping.

## Introduction:

Graphene, the new nanocarbon, has generated great scientific interest due to its unusual electronic properties [1-3] along with the possibility of controlling its conductivity by suitable doping of electrons or holes. Unlike in the case of conventional semiconductors, like Si, doping of charge carriers via heterovalent substitution has pronounced deleterious effects in graphene due to its two dimensional nature [4]. Even strongly chemisorbed ionic electron donors or acceptors, such as K [5] and $NO_2$ [6], are expected to provide strong scattering from randomly distributed donors and acceptors. Two ways to circumvent this problem has been explored so far. One, known as electrochemical doping, injects electrons or holes into graphene by the application of a suitable gate voltage with the appropriate polarity [7,8]. In the alternate, more chemical approach, conjugated organic donor or acceptor molecules, may be used as charge-transfer dopants [9]. Investigation of charge transfer dopings for a variety of adsorbed organic $e^-$ rich and deficient molecules have demonstrated that tetrathiafulvalene, TTF, and tetracyanoethylene, TCNE are effective $n$ and $p$ dopants, respectively when adsorbed on graphene (FG) [10,11]. Raman spectroscopy is being routinely used very effectively to characterize the nature and extent of doping in graphene [6,12]. The Raman G-band of graphene stiffens on both $n$ and $p$ type electrochemical doping, shifting towards higher wave numbers [7,8]. In contrast, however, $p$-type charge-transfer doping stiffens the G-band, whereas n-type doping softens it [10,11], presenting an unresolved puzzle and raising questions whether doping by the electrochemical process and that by charge-transfer molecular processes are fundamentally different. Traditionally, the process of doping has been investigated by probing changes in the doped state, namely graphene, by studying its band structure [5,6] or Raman spectra [7,8,10,11]. We turn this approach around and use x-ray photoelectron spectroscopy (XPS) to investigate the dopant states instead, since this method affords great sensitivity and allows one to quantitatively establish the occurrence of charge-transfer interaction. Using the XPS data in conjunction with Raman and electrical measurements, we are able to resolve the previously mentioned puzzle, establishing the equivalence between molecular charge-transfer doping and electrochemical doping schemes.

## Synthesis, Characterization and Methodologies:

Graphene (FG) was prepared by the exfoliation of graphite oxide in a furnace preheated to $1050^\circ$C under argon gas flow for about 30 seconds as mentioned in earlier reports [11,13,14]. The number of layers in the graphene (FG) samples is 4±1, as characterized by atomic force microscopy and analysis of the (002) reflection in the XRD pattern. The doped graphene samples were prepared by dispersing one milligram of graphene (FG) in 3 ml of benzene, containing 0.1 molar (M) of TTF or TCNE and then it was sonicated. The resulting solution was then filtered through an Anodisc filter (pore size 0.1 $\mu$m). XPS measurements were performed on as-prepared samples mounted on copper stubs with silver paste, using Al K$\alpha$ radiation (1486.6 eV) in a commercial photoelectron spectrometer from VSW Scientific Instruments. The base pressure of the chamber was maintained around $5\times10^{-10}$ mbar during the experiments. For the analysis of the XPS spectra in terms of contributions from individual components representing different species, experimental spectra were fitted by a combination of components by minimizing the total squared-error (least squared-error) of the fit. Individual components were represented by a convolution of Lorentzian function, representing the life-time broadening, and a Gaussian function to account for the instrumental resolution. The Gaussian broadening was kept the same for different components. A Shirley background-function is considered to account for the inelastic background in the XPS spectra. Raman spectra were recorded using a LabRAM HR high resolution Raman spectrometer using a He-Ne laser ($\lambda$=632.8 nm) as described earlier [11]. Electrical current vs voltage measurements were performed by drop-coating the graphene (FG) sample on Au gap electrodes patterned on glass substrates.

## Results and Discussions:

The sulfur (S) 2$p$ XPS spectrum of TTF ($n$) doped graphene (FG), shown in Fig. 1(a), shows the presence of overlapping signals from multiple sulfur species, in the doped sample. Spectral decomposition, as explained in the Methodology section, reveals the presence of three distinct S 2$p$ signals in the recorded spectrum, as shown in Fig. 1(a). In this figure, the dots represent the experimental spectrum, while thin solid lines show the three components, marked S1, S2 and S3. The sum of these three components and the background Schirley function, representing the simulated total spectrum is shown by the solid line overlapping the experimental data, showing a very good agreement. Sulfur 2$p$ signal from a wide variety of compounds [15,16,17] is known to have a simple spin-orbit doublet structure without any complication from any satellite features, making the spectral decomposition and subsequent assignment of component peaks to different species of sulfur straightforward. Binding energies (BE) of the S 2$p_{3/2}$ feature for S1, S2 and S3 sulfur atoms are 163.5 eV, 164.6 eV and 167.8 eV, respectively. The S1 species, appearing at 163.5 eV BE, is in excellent agreement with the reported S 2$p$ peak position (163.5 eV) from neutral TTF molecule [18]. The presence of the weaker S2 signal at 164.6 eV BE, which is in excellent agreement with the reported S 2$p$ peak position (164.7 eV) from positively charged TTF molecule [19], which is absent in neutral TTF molecule, clearly demonstrates that the TTF molecule indeed acts as an electron donor (*n*-doping). The observed S3 signal, shown in Fig 1(a), corresponds to the presence of sulfur atoms in a higher ($S^{6+}$) oxidation state in the sample [20,21], which is most likely generated during the preparation of the sample.

The nitrogen (N) 1$s$ spectrum of TCNE (*p*) doped graphene (FG) in Figure 1(b), also shows the presence of overlapping signals from three species of N atoms. Unlike in the case of S 2$p$ signal, it is known that shake-up satellites can complicate the spectral shape and, therefore, the analysis of N 1$s$ spectra. However, both undoped and doped TCNE exhibit single, symmetric N 1$s$ spectra without the presence of any shake-up satellites [22], thereby allowing an unambiguous spectral decomposition, enabling us to assign various features of N 1$s$ spectrum in Figure 1(b) to distinct species of N. Similar spectral decomposition, as explained in the Methodology section and used for S 2$p$ spectral analysis, reveals the presence of three distinct N 1$s$ signal in the recorded spectrum, as shown in Fig 1(b). Similarly in Fig 1(b), the dots represent the experimental spectrum, the thin solid lines represent the individual components and the thick solid line represents the combined simulated spectrum, which shows a very good agreement with the experimental spectrum. The species, at ~400.4 eV BE, denoted by N1, is in excellent agreement with the reported N 1$s$ peak position (400.4 eV) from neutral TCNE molecule [22]. The stronger nitrogen signal, denoted by N2, at a lower BE of ~398.7 eV, is in excellent agreement with the reported peak position of N 1s (398.8 eV) signal from negatively charged TCNE [22]. Furthermore, widths of individual species of N1 and N2 (1.8 and 1.6 eV respectively) are also in good agreement with those (~ 2.0 and 1.7 eV, as estimated from the reported spectra) for pure species of neutral and negatively charged TCNE, respectively [22], endorsing the assignment of these peaks. N3 signal, occurring at a BE of 401.9 eV, corresponds to the presence of oxidized nitrogen cations. We note that the presence of radical cations of sulphur and nitrogen has also been observed in the electronic absorption spectra of TTF doped and TCNE doped graphene respectively [11]. The presence of charge-transfer related N2 species of nitrogen with a negative chemical shift clearly establishes that TCNE acts as a $p$-type dopant, receiving electrons from graphene. Further the enhanced intensity of the negatively charged TCNE species (N2), compared to the signal from neutral TCNE (N1) in TCNE doped graphene (FG), as opposed to the case for TTF doped graphene (FG), for the same concentration of dopant molecules present, further shows that $p$-doping in graphene is stronger compared to $n$-doping. This is not surprising, given the fact that undoped graphene is electron rich due to the delocalized $\pi$ ($e^-$) charge cloud on it. This observation is consistent with the higher interaction energy of TCNE with graphene (FG) compared to TTF as measured calorimetrically [23].

Having qualitatively established charge transfer dopings from TTF (n) and TCNE (p), we obtain quantitative estimate of the extent of charge transfer in TTF doped graphene (FG) by calculating the ratio of the areas under the S2 sulfur species, which corresponds to charged TTF molecules (which have donated electrons to graphene) to that for C 1s of graphene (after subtracting the contributions of carbon atoms from TTF molecules) and correcting for different cross-sections of S 2$p$ and C 1$s$ for Al K$\alpha$ radiation. Assuming a homogeneous distribution of dopant molecules, the doped electron ($n$) concentration for 0.1 M TTF doped graphene (FG) is found to be $0.6*10^{12}$ cm$^{-2}$. A quantitative estimation of doped hole concentration for TCNE-doped graphene (FG) was performed similarly using the area under the N2 nitrogen species and C1$s$ and correcting for the cross-sectional effects; the doped hole ($p$) concentration for 0.1 M TCNE doped graphene (FG) is found to be $0.4*10^{13}$ cm$^{-2}$. Thus, we find that an order of magnitude higher doping can be achieved in the case of $p$-doping with TCNE compared to $n$-doping with TTF on graphene.

Fig. 2 shows the C 1$s$ spectrum from undoped graphene along with those from graphene doped with TCNE and TTF. The C 1$s$ peak of graphene matches with the sp$^2$ hybridized graphitic carbon [24]. All three spectra overlap on each other within the experimental uncertainty. This indicates that the shift of the C 1$s$ signal for electron and hole dopings in the present case remain below the detection limit. This is consistent with the doping level ($\sim 10^{12}$ cm$^{-2}$) deduced here, as such a small doping level is not expected to have any perceptible effect on the binding energy of the C 1$s$ peak. Similarly, no shift in the valence band edge near the Fermi energy could be found in valence band spectra of these compounds (see Fig. 3).

Raman spectra recorded for various concentrations of TTF ($n$) and TCNE ($p$) in charge transfer doped graphene (FG) sample, show that the position of the Raman G-band, as shown in Fig. 4(a), stiffens progressively with increasing concentration of the p-dopant, TCNE. This is consistent with the results of stiffening of Raman G-band by p-type doping, achieved by applying negative top/back gate voltages on graphene in electrochemical doping [7,8]. However, the Raman G-band softens progressively with the concentration of the n-type charge transfer dopant, TTF (Fig. 4(a)). This is in contradiction to the result from electrochemical $n$ doping, achieved by applying positive top/back gate voltages on graphene, where similar to p-type doping, the Raman G-band continues to stiffen for increasing $n$ concentrations [7,8]. This apparent anomaly can, however, be resolved by noting that the undoped graphene sample is intrinsically hole-doped, as also observed in most graphene samples. For such an intrinsically hole doped graphene sample, a certain threshold of doped electron concentration is needed for the Fermi energy ($E_F$) to cross the Dirac point and, consequently, in order for the charge carriers to become electron-like. Below this threshold concentration, the carriers continue to remain hole-like, in spite of the electron doping which is essentially used up in compensating the initial, intrinsic doped holes in such graphene samples. Since the Raman G-band is known to stiffen for both n and p type doping, away from the Dirac point, it now becomes easy to understand that for an intrinsically hole doped graphene sample, *n* type doping upto the threshold electron concentration will only soften the Raman G-band, as electron doping at this level represents an approach of the Fermi level towards the Dirac point and not away from it. The doped electron concentration in the present case ($0.6*10^{12}$ cm$^{-2}$ for 0.1 M TTF) is less than the threshold concentration, estimated[8] to be $5*10^{12}$ cm$^{-2}$, which enables us to understand the reason behind the softening of Raman G-band with increasing TTF concentrations. This conclusion is supported by the observation that the conductance of the doped sample, obtained by taking d(current (I))/d(applied voltage(V)) at V = 0 Volts (Fig. 4(b)), decreases continuously with increasing TTF (*n*) concentration in the sample. This directly correlates with the diminishing number of hole carriers on increasing the doped electron concentration, brought about by additional TTF molecules. Thus, we find that the Fermi level for even the highest concentration (0.1 M) of TTF doped graphene (FG) sample still lies below the Dirac-like point and charge carriers are still hole-like, as shown schematically in Figure 5. The conductance, as shown in Fig. 4(b), however, continues to increases with increasing *p* dopant concentration brought about by addition of TCNE, as the number of hole carriers increases.

In conclusion, we have established the occurrence of charge-transfer in graphene covered with TTF and TCNE. We are able to quantitatively understand the extent of charge transfer doping in each case. Combining XPS, Raman and electrical measurements, we have provided a coherent understanding of doping through charge transfer and electrochemical doping.

Authors thank Department of Science and Technology, Government of India for funding and Prof. A. K. Sood for useful discussions.

**References:**

*Also at JNCASR, Bangalore-560054, India.

Electronic mail: sarma@sscu.iisc.ernet.in

Figure Captions:

Figure 1: (a) S 2$p$ core level spectrum of TTF-doped graphene (FG), showing the presence of three species of sulfur in the sample, S1, S2 and S3 corresponding to neutral TTF, positively charged TTF and oxidized sulfur species. (b) N 1$s$ core level spectrum of TCNE doped graphene (FG), showing the presence of three species of nitrogen in the sample, NI, N2 and N3 corresponding to neutral TCNE, negatively charged TCNE and oxidized nitrogen species respectively. Open circles corresponds to the experimental data and the solid line running through, denotes the theoretical fit in both the panels.

Figure 2: The C1$s$ core level spectra for undoped, TCNE ($p$) doped and TTF ($n$) doped graphene, showing the absence of any chemical shifts. The inset shows an expanded view for better comparison.

Figure 3: The near Fermi edge (BE=0 eV) spectra of undoped, TCNE ($p$) doped and TTF ($n$) doped graphene.

Figure 4: (a) Position of Raman G-band of doped graphene (FG) sample for different concentrations of TCNE ($p$) dopant and TTF ($n$) dopant molecules. [Adapted from Ref. 11]
(b) The dependence of conductance (dI/dV) of doped graphene (FG) samples on the concentration of TCNE and TTF dopant molecules.

Figure 5: A schematic view of the position of Fermi level with respect to the Dirac point for 0.1 M TTF ($n$) doped, intrinsic and 0.1 M TCNE ($p$) doped graphene. The schematic shows that graphene samples are intrinsically  hole ($p$) doped. The charge carriers for 0.1 M TTF ($n$) doped graphene still continues to remain hole-like, explaining the unusual shift of Raman G-band.

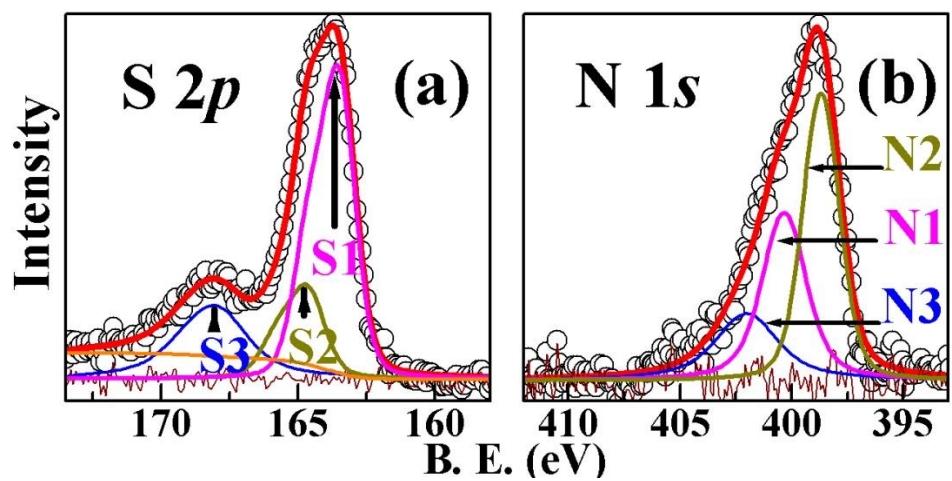

Figure 1

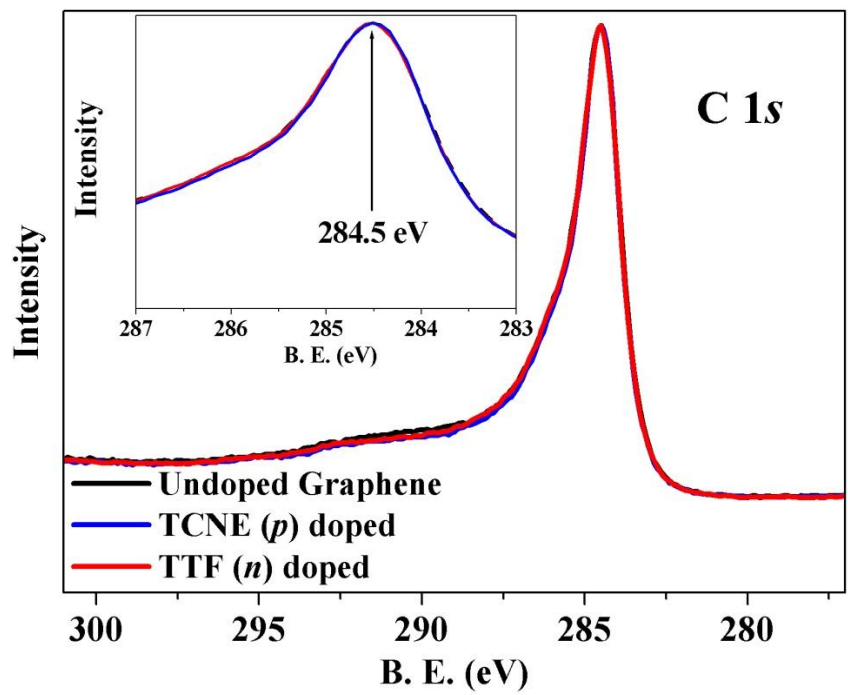

Figure 2

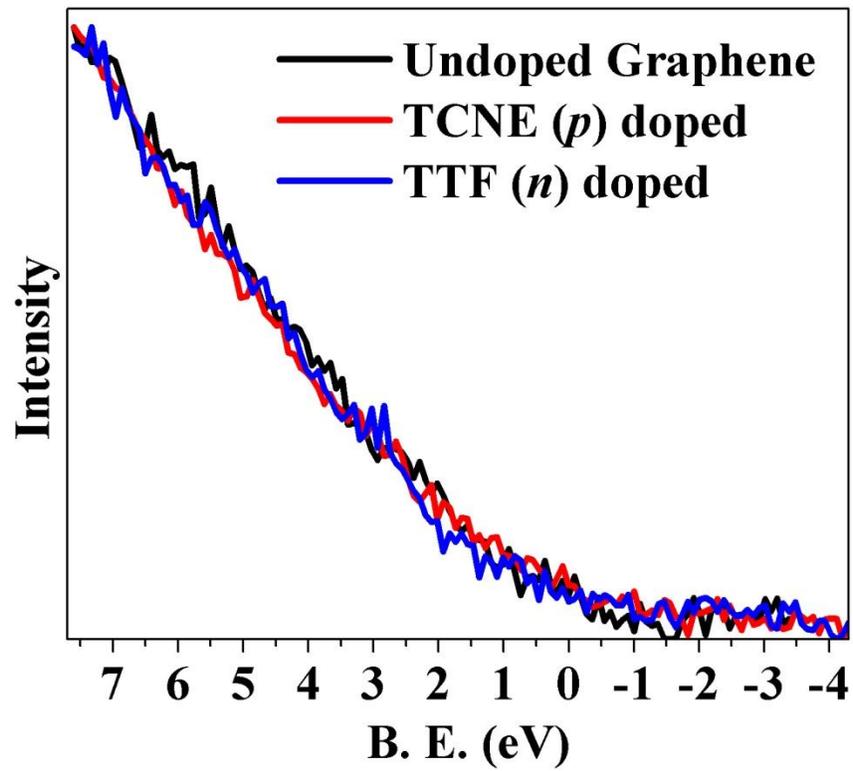

Figure 3

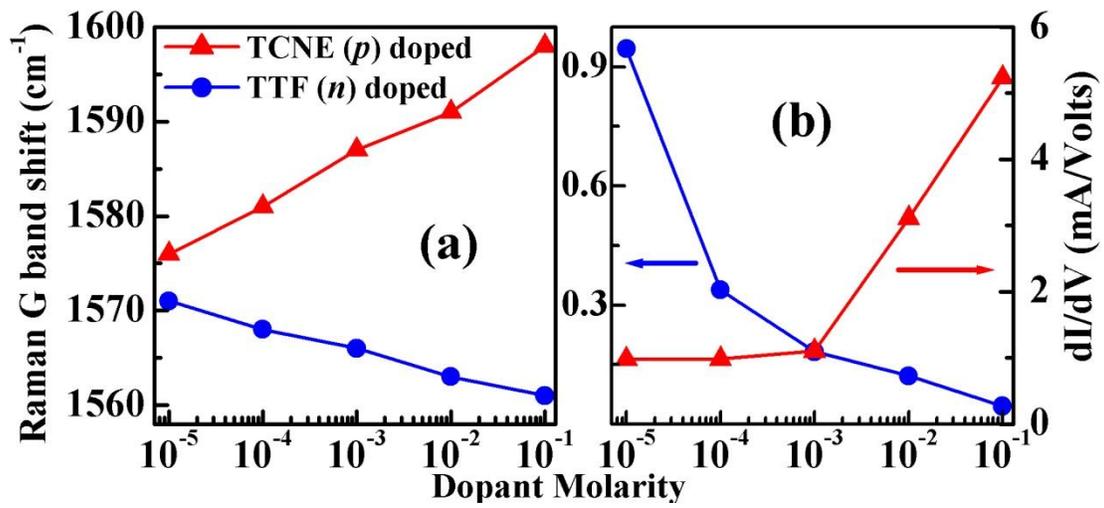

Figure 4

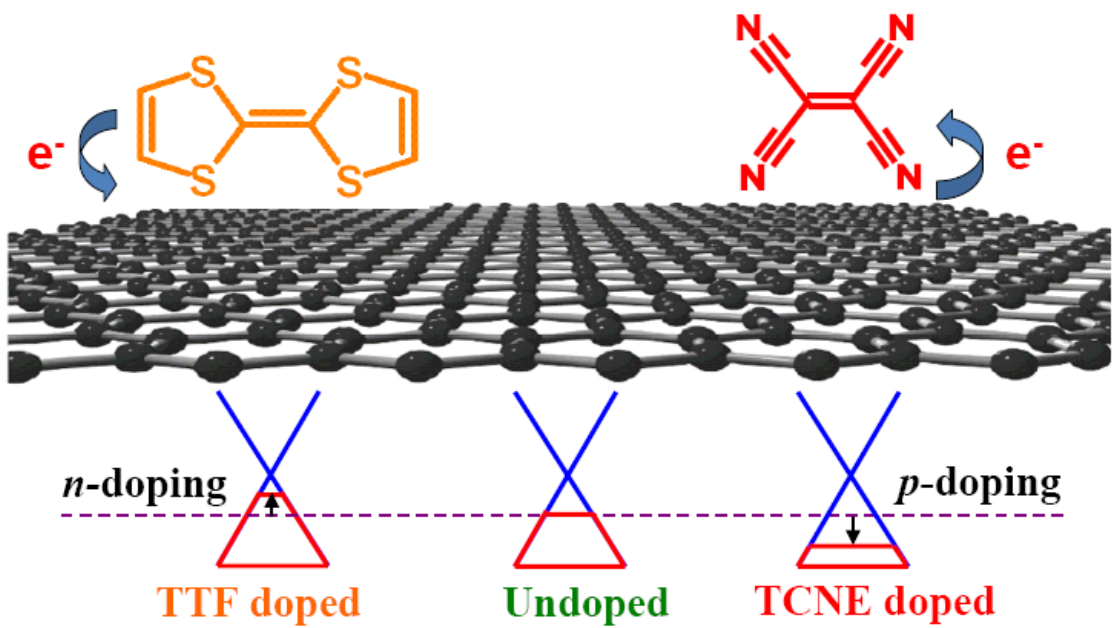

Figure 5